\documentstyle[preprint,aps]{revtex}

\def\ut#1{\rlap{\lower1ex\hbox{$\sim$}}{#1}}
\def\l{\ell_{P}}

\def\SU{{\rm SU}}
\def\SL{{\rm SL}}
\def\SO{{\rm SO}}
\def\D{{\cal D}}
\def\R{{\rm I}\!{\rm R}}
\def\E{{}^\gamma\!\Sigma}
\def\ET{{}^\gamma\!\tilde{E}}
\def\A{{}^\gamma\!A}

\begin{document}
\draft
\title{On the Constant that Fixes the Area Spectrum in \\
Canonical Quantum Gravity}
\author{K.\ Krasnov\thanks{E-mail address: krasnov@phys.psu.edu}}
\address{Center for Gravitational Physics and Geometry, \\
The Pennsylvania State University, PA 16802, USA.}

\maketitle

\begin{abstract}
The formula for the area eigenvalues that was obtained by
many authors within the approach known as loop quantum gravity
states that each edge of a spin network contributes an area 
proportional to $\sqrt{j(j+1)}$ times Planck length squared
to any surface it transversely intersects. However, some 
confusion exists in the literature as to a value of the
proportionality coefficient. The purpose of this rather technical 
note is to fix this coefficient. We
present a calculation which shows that in a sector of
quantum theory based on the connection $A=\Gamma-\gamma K$,
where $\Gamma$ is the spin connection compatible with the triad field,
$K$ is the extrinsic curvature and $\gamma$ is Immirzi parameter,
the value of the multiplicative factor is $8\pi\gamma$. In other words,
each edge of a spin network contributes an area $8\pi\gamma\l^2\sqrt{j(j+1)}$
to any surface it transversely intersects. 
\end{abstract}
\pacs{ }

One of the most intriguing results of canonical quantum gravity based
on Ashtekar's new variables \cite{NewVar} and the loop representation
of Rovelli and Smolin \cite{RS} is the discreteness of the area spectrum
\cite{Area}.  In this theory, there is
Hilbert space of kinematical states with a basis given by spin networks,
i.e., graphs in space with edges labelled by spins $j = {1\over 2},1,
\dots$ and vertices labelled by intertwining operators \cite{Network}.
Each edge contributes an area proportional to $\sqrt{j(j+1)}$ times
Planck length $\l$ squared to any surface it transversely intersects.  
It was also realized by Immirzi \cite{ImmPar} that an additional
parameter enters the formula for the area spectrum due to
a quantization ambiguity. Namely, instead of working with a complex
connection $A_a = \Gamma_a - i K_a$, where $\Gamma_a$ is the
3-dimensional spin connection compatible with the triad field 
and $K_a$ is the extrinsic curvature of the spatial hypersurface,
one can introduce real phase space variables of Barbero \cite{Real-var}: 
\begin{eqnarray}
\A_a := \Gamma_a - \gamma K_a \nonumber \\
\E_{ab} := (1/\gamma) \Sigma_{ab},
\label{realvar} 
\end{eqnarray}
where $\gamma > 0$ is real and is known as Immirzi parameter,
and $\Sigma_{ab}$ is the two-form dual to the densitized triad field.
Then there is a one-parameter family
of inequivalent quantizations depending on which of real $\SU(2)$
connections $\A_a$ have
been used to construct the quantum theory (see Rovelli and Thiemann
\cite{ImmPar}). The parameter $\gamma$ explicitly enters the 
formula for the area eigenvalues: each transverse intersection
contributes an area proportional to $\gamma\l^2\sqrt{j(j+1)}$ \cite{ImmPar}.

The purpose of this note is to find the value of the proportionality
coefficient. To find this coefficient we start from 
the self-dual action \cite{Action} adjusting the multiplicative
constant in front of it so that it coincides with 
Einstein-Hilbert action when the self-dual connection 
satisfies its equations of motion. One can then find Hamiltonian
formulation of the theory, and perform a canonical transformation
to the real variables (\ref{realvar}). Knowing the
simplectic structure on the phase space $(\A,\E)$
one can use one of the methods \cite{Area} to calculate the area
spectrum. Our result is that the proportionality constant is
equal to $8\pi$, i.e., the area spectrum is 
\begin{equation}
8\pi\gamma\l^2\sum_p\sqrt{j_p(j_p+1)},
\label{qarea}
\end{equation}
where the sum is taken over all points $p$ on the surface where edges of 
a spin network state intersects this surface, $j_p$ are spins (half-integers) that 
label the corresponding edges, $\gamma$ is the real parameter as in (\ref{realvar}),
and $\l^2=G\hbar$, $G$ being Newton constant.

We are aware of another attempt (see De Pietri and Rovelli \cite{Area}) 
to fix the value of the multiplicative
factor in (\ref{qarea}). However, Eq. (2.4) of their paper should
read 
\begin{equation}
{1\over 2G_{DR}} \int d^4x \sqrt{-g} R =
{1\over G_{DR}} \int dx^0 \int d^3x \,\bigl[\cdots\bigr],
\end{equation}
and as a consequence the constant $16\pi G_{DR}$ that the authors call Newton's constant
is 2 times less the one usually called Newton's constant.

We start from the self-dual formulation of general relativity \cite{Action}.
The action 
\begin{eqnarray}
\label{action}
S[\sigma,A]=-{i\over 8\pi G}\int_{\cal M}\,{\rm Tr}\left(\Sigma\wedge F\right)= \\
{i\over 8\pi G}\,{1\over4}\,\int_{\cal M} d^4x \,\Sigma_{ab}^{AB}\,F_{cd AB}
\tilde{\varepsilon}^{abcd}
\nonumber
\end{eqnarray}
is a functional of tetrad $\sigma_a^{AA'}$ and the self-dual connection $A_a^{AB}$
fields. Here $\cal M$ is the spacetime manifold (which we for simplicity 
assume to be of the topology $\R\times M$ where $M$ is some compact
manifold), primed and unprimed upper case letters stand for $\SL(2,C)$ spinor 
indices, lower case letters denote spacetime indices,
$\tilde{\varepsilon}^{abcd}$ is Levi-Civita density, 
and the tetrad field defines the metric via
\begin{equation} \label{metric}
g_{ab} = \sigma_a^{AA'}\sigma_{b AA'}.
\end{equation}
The tetrad field $\sigma$ is required to be anti-hermitian 
($\overline{\sigma}_a^{AA'} = -\sigma_a^{A'A}$) so that the metric
(\ref{metric}) is real Lorentzian metric of signature
$(-+++)$. The self-dual connection $A$ defines a derivative
operator $\D_a$ that operates on unprimed spinors 
$\D_a\lambda_A=\partial_a\lambda_A + A_{a A}^{\,\,\,\,\,B}\lambda_B$. The
field $\Sigma$ in (\ref{action}) is the two-form
$\Sigma^{AB}:=\sigma^{AA'}\wedge\sigma^B_{\,A'}$, or, in 
index notations $\Sigma_{ab}^{AB}=2\sigma_{[a}^{AA'}\sigma^B_{b]\,A'}$. 
Note that $\Sigma$ is self-dual also in spatial indices
\begin{equation}
{1\over2}\varepsilon^{abcd}\Sigma_{ab}^{AB}=i\,\Sigma^{cd\,AB},
\label{self-dual}
\end{equation}
where $\varepsilon$ is the natural volume 4-form defined by the
metric. The field $F$ in (\ref{action}) is the curvature two-form 
of the connection $A$: $2\D_{[a}\D_{b]}\lambda_A=F_{ab A}^{\,\,\,\,\,\,B}\lambda_B$.

We want to prove now that the action (\ref{action}) is equal to 
the Einstein-Hilbert action
\begin{equation}
{1\over16\pi G} \int_{\cal M} \sqrt{-g} R
\end{equation}
when the connection $A$ satisfies its equations of motion. Varying
(\ref{action}) with respect to $A$ one gets 
(when $\sigma$ is non-degenerate) $\D_a\sigma_b^{AA'}=0$. This
equation gives one a relation between curvature $F$ and the 
Riemann curvature tensor: $F$ turns out to be the self-dual part
of the Riemann tensor (see \cite{Book}, p. 292)
\begin{equation}
R_{abc}^d = F_{ab B}^A\sigma_c^{BA'}\sigma^d_{AA'} +
\overline{F}_{ab B'}^{A'}\sigma_c^{AB'}\sigma^d_{AA'}.
\end{equation}
Now, using self-duality (\ref{self-dual}) of $\Sigma$ one can
show that
\begin{equation} \label{1}
F_{ab}^{AB} = - {1\over 4}{}^+\!R_{abcd}\Sigma^{cd\,AB} =
- {1\over 4} R_{abcd}\Sigma^{cd\,AB},
\end{equation}
where ${}^+\!R_{abcd}$ is the self-dual part of Riemann tensor and 
self-duality of $\Sigma$ was used to get the
second identity.

Thus, when the connection satisfies its equations of motion the
action (\ref{action}) is equal to 
\begin{eqnarray}
{i\over 8\pi G}\int_{\cal M} d^4x\, i\sqrt{-g}\,\Sigma^{cd AB} {1\over2}F_{cd AB} = \\
\nonumber
{i\over 8\pi G}\int_{\cal M} d^4x\, i\sqrt{-g}\,\Sigma^{cd AB} \left(-{1\over8}\right)
R_{cdef}\Sigma^{ef}_{AB},
\end{eqnarray}
where self-duality of $\Sigma$ was used in the first line, and 
(\ref{1}) in the second. Now, using 
identity \cite{CDJ}
\begin{equation}
\Sigma^{ab AB}\Sigma^{cd}_{AB} = 4 g^{a[c}g^{d]b} - 2i\epsilon^{abcd},
\end{equation}
we have for the action
\begin{equation}
{1\over 16\pi G}\int_{\cal M} d^4x \sqrt{-g} R_{cdef} g^{a[c}g^{d]b} =
{1\over 16\pi G}\int_{\cal M} d^4x \sqrt{-g} R,
\end{equation}
where the kinematical Bianchi identity $R_{[abc]}^{\,\,\,\,\,\,\,\,d}=0$ was used to get the
first formula. This proves that the action (\ref{action}) is 
equivalent to Einstein-Hilbert action.

We now want to find a phase space formulation of the theory and perform
a canonical transformation to the real variables (\ref{realvar}). We
use the covariant description in which the phase space is the space
of solutions of the classical equations of motion. The simplectic
structure on the phase space is determined by the 3-form
$\bf \Theta$ that is found by varying the lagrangian $\bf L$ of the 
theory $\delta {\bf L} = {\bf E}\delta\phi + d{\bf\Theta}(\phi,\delta\phi)$,
where $\phi$ denotes all of the dynamical fields and $\bf E$
is the 4-form giving the equations of motion 
(see, for example, \cite{Wald}). For the case of theory 
defined by (\ref{action}) we have
\begin{equation}
{\bf\Theta} = -{i\over 8\pi G} {\rm Tr}\left(\Sigma\wedge\delta A\right).
\end{equation}
Thus, the simplectic structure on the phase space is
\begin{equation}
\Omega|_{(A,\Sigma)} 
\left ( (\delta A,\delta\Sigma), (\delta A',\delta\Sigma') \right ) =
-{i\over 8\pi G} \int_M {\rm Tr}
\left [ \delta\Sigma\wedge\delta A' - \delta\Sigma'\wedge\delta A \right ],
\end{equation}
where the integral is taken over a Cauchy surface $M$. As usual, it
does not depend on a choice of $M$. The points of phase
space are labelled by restrictions of fields $A,\Sigma$ on $M$.

On $M$, $A$ can be expressed in terms of real
fields as $A = \Gamma - i K$, where $\Gamma_a$ is the
3-dimensional spin connection compatible with the triad 
(the densitized triad vector field is dual to the two form $\Sigma$)
and $K$ is the extrinsic curvature of $M$. This suggests \cite{ImmPar} that
we introduce real phase space variables (\ref{realvar}). Since
$M$ is compact, the canonical transformation to variables
(\ref{realvar}) is well-defined, and
one can easily check that in terms of these manifestly real variables,
the simplectic structure on the phase space is given by
\begin{equation}
\Omega|_{(\A,\E)} 
\left ( (\delta\A,\delta\E), (\delta\A',\delta\E') \right ) =
{1\over 8\pi G} \int_M {\rm Tr}
\left [ \delta\E\wedge\delta\A' - \delta\E'\wedge\delta\A \right ].
\end{equation}

To compare the simplectic structure found with the one
that was the starting point in the calculation of the 
area spectrum in \cite{Area},
we now convert all $\SU(2)$ indices to $\SO(3)$ indices via
\begin{eqnarray}
\A_{a A}^{\,B}:= -{i\over2}\tau_A^{i\,B} \A_a^i \\
\ET_A^{a\,B}:= -{i\over\sqrt{2}}\tau_A^{i\,B} \ET^{ai},
\nonumber
\end{eqnarray}
which are the standard conventions in the literature (see, for example,
\cite{Book}). Here $\tau_A^{i\,B}$ are Pauli matrices 
$(\tau^i\tau^j)_A^{\,B}=i\varepsilon^{ijk}\tau_A^{k\,B} +
\delta^{ij}\delta_A^{\,B}$, and
the densitized triad field $\ET$ is related
to $\E$ via 
\begin{equation}
\E_{ab}^{AB} = \sqrt{2}\ut{\varepsilon}_{abc}\ET_A^{c\,B}.
\end{equation}
To get the last formula the relation \cite{Book}, Appendix A (43') was used.
Using these standard definitions and identities one can 
find the simplectic structure to be equal to
\begin{equation}
\Omega|_{(\A,\ET)} 
\left ( (\delta\A,\delta\ET), (\delta\A',\delta\ET') \right ) =
-{1\over8\pi G}\int_M d^3x 
\left[ \delta\ET^{ai}\delta{\A'}_{ai} - \delta{\ET'}^{ai}\delta\A_{ai} \right ].
\end{equation}
Thus, the Poisson brackets between the canonical variables $(\A,\ET)$
are given by
\begin{equation} \label{2}
\left\{ \A_{ai}(x), \ET^{bj}(y) \right\} =
8\pi G \delta_i^j\delta_a^b\tilde{\delta}(x,y).
\end{equation}

The Poisson brackets in the form (\ref{2}) is the starting point of
the calculation of the area spectrum performed by Ashtekar and 
Lewandowski \cite{Area}. Comparing (\ref{2}) with the formula
(2.1) from that paper, we find $G_{AL}=8\pi G$, where $G_{AL}$
is the the constant of the dimension of Newton's constant used
by Ashtekar and Lewandowski, and $G$ is Newton's constant. We
refer now to the calculation performed in that paper to 
fix the area spectrum to be (\ref{qarea}).

{\bf Acknowledgements}: I am grateful to A. Ashtekar and R. De Pietri
for discussions. This work was supported in part by the NSF
grant PHY95-14240 and by the Eberly research funds of Penn State.


\begin{thebibliography}{99}
\bibitem{NewVar} A.\ Ashtekar, Phys.\ Rev.\ Lett.\ {\bf 57}, 2244 (1986);
Phys.\ Rev.\ {\bf D36}, 1587 (1987).

\bibitem{RS} C.\ Rovelli and L.\ Smolin, Nucl.\ Phys.\ {\bf B331}, 80
(1990).  

\bibitem{Area} C.\ Rovelli and L.\ Smolin, Nucl.\ Phys.\ {\bf B442},
593 (1995); R.\ De Pietri and C.\ Rovelli, Phys.\ Rev.\ {\bf D54},
2664 (1996); S.\ Fritelli, L.\ Lehner, C.\ Rovelli, Class.\ Quant.\
Grav.\ {\bf 13}, 2921 (1996); A.\ Ashtekar, J.\ Lewandowski, Class.\ 
Quant.\ Grav.\ {\bf 14}, 55 (1997).

\bibitem{Network} A.\ Ashtekar and J.\ Lewandowski, in {\sl Knots and Quantum
Gravity}, edited by J.\ Baez (Oxford U.P., Oxford, 1994); J.\ Baez,
Lett.\ Math.\ Phys.\ {\bf 31}, 213 (1994); C.\ Rovelli and L.\ Smolin,
{\sl Phys.\ Rev.\ }{\bf D52}, 5743 (1995); J.\ Baez, 
{\sl Adv.\ Math.\ }{\bf 117}, 253 (1996).

\bibitem{ImmPar}  G.\ Immirzi, ``Quantum Gravity and Regge
Calculus'', gr-qc/9701052; C.\ Rovelli, T.\ Thiemann, 
``The Immirzi parameter in quantum general relativity'', gr-qc/9705059.

\bibitem{Real-var} A. Ashtekar, in {\it Mathematics and General Relativity},
edited by J. Isenberg, (AMS, Providence 1989); F. Barbero, Phys.\
Rev.{\bf D54}, 1492 (1996); T.\ Thiemann, Class.\ and Quant.\
Grav.\ {\bf 13}, 1383 (1996).

\bibitem{Action} J.\ F.\ Plebanski, J.\ Math.\ Phys.\ {\bf 18}, 2511 (1977);
T.\ Jacobson and L.\ Smolin, Nucl.\ Phys.\ {\bf B299}, 583 (1988).

\bibitem{Book} A.\ Ashtekar, Lectures on non-perturbative canonical
gravity, World Scientific Publishing Co., 1991.

\bibitem{CDJ} R.\ Capovilla, J.\ Dell, T.\ Jacobson, L.\ Mason,
Class.\ Quant.\  Grav.\ {\bf 8}, 41 (1991).

\bibitem{Wald} V.\ Iyer, R.\ Wald, Phys.\ Rev.\ {\bf D50}, 846 (1994).

\end{thebibliography}
\end{document}